\def\heii{He {\sc ii}}
\def\hi{H~{\sc i}}
\def\hii{H~{\sc ii}}

\def\msyr{~$M_{\odot}$~yr$^{-1}$}
\newcommand{\ms}{$M_{\odot}$}

\documentclass[structabstract]{aa}
\usepackage{graphicx}
\usepackage{rotating}
\usepackage{txfonts}
\usepackage{longtable,lscape,color}

\begin{document}

\title{The chemical evolution of IC10}
\author{J.~Yin\inst{1,2}, L.~Magrini \inst{3}, F.~Matteucci \inst{2,4},
G.~A.~Lanfranchi \inst{5}, D.~R.~Gon\c calves \inst{6},
R.~D.~D.~~Costa \inst{7}}


\institute{ Key Laboratory for Research in Galaxies and Cosmology,
Shanghai Astronomical Observatory, Chinese Academy of Sciences, 80
Nandan Road, Shanghai, 200030, China.  \email{jyin@shao.ac.cn} \and
Dipartimento di Fisica, Sezione di Astronomia, Universit\`a di
Trieste, via G.B. Tiepolo 11, 34131 Trieste, Italy.
\email{yin--matteucc@oats.inaf.it} \and INAF -- Osservatorio
Astrofisico di Arcetri, Largo E. Fermi, 5, I-50125 Firenze, Italy.
\email{laura@arcetri.astro.it} \and INAF -- Osservatorio Astronomico
di Trieste, via G.B. Tiepolo 11, 34131 Trieste, Italy. \and NAT --
Universidade Cruzeiro do Sul, R. Galv\~ao Bueno 868, 01506-000,
S\~ao Paulo, Brazil. \email{gustavo.lanfranchi@cruzeirodosul.edu.br}
\and UFRJ -- Observat\'orio do Valongo, Ladeira Pedro Antonio 43,
20080-090 Rio de Janeiro, Brazil. \email{denise@astro.ufrj.br} \and
IAG -- Universidade de S\~ao Paulo, Rua do Mat\~ao 1226, 05508-900
S\~ao Paulo, Brazil. \email{roberto@astro.iag.usp.br} }

\date{}
\abstract {Dwarf irregular galaxies are relatively simple unevolved
objects where it is easy to test models of galactic chemical
evolution.} {We attempt to determine the star formation and gas
accretion history of IC10, a local dwarf irregular for which
abundance, gas, and mass determinations are available.} {We apply
detailed chemical evolution models to predict the evolution of
several chemical elements (He, O, N, S) and compared our predictions
with the observational data. We consider additional constraints such
as the present-time gas fraction, the star formation rate (SFR), and
the total estimated mass of IC10. We assume a dark matter halo for
this galaxy and study the development of a galactic wind. We
consider different star formation regimes: bursting and continuous.
We explore different wind situations: {\em i}) normal wind, where
all the gas is lost at the same rate and {\em ii}) metal-enhanced
wind, where metals produced by supernovae are preferentially lost.
We study a case without wind. We vary the star formation efficiency
(SFE), the wind efficiency, and the time scale of the gas infall,
which are the most important parameters in our models.}{We find that
only models with metal-enhanced galactic winds can reproduce the
properties of IC10. The star formation must have proceeded in bursts
rather than continuously and the bursts must have been less numerous
than $\sim$10 over the whole galactic lifetime. Finally, IC10 must
have formed by a slow process of gas accretion with a timescale of
the order of 8 Gyr.}{}

\keywords{Galaxies: abundances - Galaxies: evolution - Galaxies:
irregular - Galaxies: dwarf - Galaxies: ISM - Galaxies, individual:
IC10}
\authorrunning{J. Yin et al.}
\titlerunning{The chemical evolution of IC10}
\maketitle

\section{Introduction}

Understanding how dwarf galaxies formed is a key aspect of modern
observational cosmology because they are currently the most abundant
galaxy population, and were probably even more abundant during the
first epochs of the Universe. In addition, according to the
hierarchical formation scenario (e.g., White \& Frenk
\cite{white91}), they are building blocks of larger structures,
hence their study is crucial to understanding galaxy formation and
evolution processes. Several questions about their formation and
evolution still remain unsolved, among them: {\em i)} the origin of
the differences between the irregular and spheroidal morphological
types (e.g., Mateo \cite{mateo98}), {\em ii)} the relation between
their mass and metallicity, {\em iii)} the contribution of galactic
and supernovae (SNe) winds to the evolution of the interstellar
medium (ISM) metallicity, and {\em iv)} their age and star formation
history (SFH) (cf., e.g., Gavil\'an et al. \cite{gavilan09}). Local
Universe dwarf galaxies are unique environments for studying these
aspects. Owing to the intrinsic faintness of dwarf galaxies,
observational constraints to their evolution, such as the chemical
abundances of stellar populations of different ages, can be obtained
only for nearby galaxies.

Therefore, the large amount of information available for Local Group
(LG) dwarf galaxies, such as star formation rate (SFR),  \hi\ and
H$_2$ contents, star surface density, and metallicity distribution,
allows us to use them to test chemical evolution models of different
kind of galaxies (cf. Carigi et al. \cite{carigi02}; Lanfranchi \&
Matteucci \cite{lanfranchi03}; Moll\'a \& D\'\i az \cite{molla05}).
Elemental abundance ratios and their variation over cosmic time
because of ongoing star formation (SF) are among the most important
constraints of chemical evolution models (e.g., Moll\'a et al.
\cite{molla96}). In this framework, optical spectroscopy of
emission-line gas in galaxies is essential for deriving chemical
abundances. In particular,  \hii~regions and planetary nebulae (PNe)
provide information about the gas chemical composition at two
different epochs in the galactic history: \hii~regions provide the
present time chemical abundances, while PNe probe the distant past
chemistry.

Dwarf irregular galaxies (dIrrs) are of great interest because they
enable detailed studies of issues such as the occurrence of galactic
winds and the chemical enrichment of the interstellar and
intergalactic media by the direct measurement of the metal content
of their ISM. In addition, their low level of evolution, as implied
by their low metallicity and high gas content, makes these systems
the most similar to primeval galaxies and, therefore, the most
useful for inferring the primordial galaxy conditions. It has also
been proposed that dwarf irregulars could represent the local
counterparts of faint blue galaxies found in excess in deep galaxy
counts (e.g., Ellis \cite{ellis97}).

In this work, we deal with the detailed chemical evolution of the
dwarf irregular galaxy IC10, building several chemical evolution
models constrained by the observed properties of this galaxy. Our
choice is motivated by recent observations (e.g., Sanna et
al.\cite{sanna09}; Magrini \& Gon\c calves \cite{magrini09},
hereafter MG09) that place strong constraints on both its SFH and
its chemical content and enrichment.

We describe the baryonic content, the SFR, and the chemical
abundances of IC10 in Sect.~2. The framework of the modelling is
given in Sect.~3, whereas the results are presented in Sect.~4.
Finally, Sect.~5 is devoted to the conclusions.

\section{Observational constraints}
IC10 is a low-mass, metal-poor, and actively star-forming galaxy. It
was first recognized to be an extragalactic object by Hubble
(\cite{hubble36}), who defined it as \lq one of the most curious
objects in the sky'. It is a member of  the LG, possibly belonging
to the Andromeda subgroup. Its location at low Galactic latitude
implies quite uncertain determinations of its reddening (E(B-V) from
0.47 to 2.0) and distance (from 0.5 to 3~Mpc, see, e.g., Kim et al.
\cite{kim09} for a summary of distances and reddenings). Distance
estimates locate IC10 at around $0.7-0.8$ Mpc (e.g., Demers et al.
\cite{demers04}; Kniazev et al. \cite{kniazev08}; Sanna et al.
\cite{sanna08}; Kim et al. \cite{kim09}).

\begin{table}
\caption{Properties of IC10 }\label{Tab:obs}
\begin{center}
\begin{tabular}{lll}
\hline
\hline
Hubble type          &  Irr~IV                                 & 1\\
Distance             & 740 kpc                                 & 2\\
Dynamical mass       & $\sim1.7\times10^{9}~M_{\odot}$         & 3 \\
Stellar mass         & $\sim(4-6)\times10^{8}~M_{\odot}$         & 4, 5, 6 \\
\hi\ mass            & $\sim(1.7-2.3)\times10^{8}~M_{\odot}$     & 6, 7, 8, 9, 10 \\
H$_2$ mass           & $1\times10^7~M_{\odot}$                 & 1\\
gas fraction         & $\sim0.22-0.36$                         &  \\
$\log$(O/H) + 12     & $8.2-8.3$                               &11, 12, 13, 14, 15\\
$[$Fe/H$]_{\rm RGB}$ & $-1.1$                                  &16             \\
SFR$_{\rm H\alpha }$ & $0.2~M_{\odot }$ yr$^{-1}$              &7, 17\\
SFR$_{\rm FIR}$      & $0.05~M_{\odot }$ yr$^{-1}$             &18, 19\\
SFR$_{1.4~\rm GHz}$  &$0.07~M_{\odot }$ yr$^{-1}$              &19, 20\\
SFR$_{\rm PNe}$      & $0.02-0.04~M_{\odot }$ yr$^{-1}$        &14\\
\hline
\end{tabular}\\
\tablebib{ (1) van den Bergh (\cite{vdb00}); (2) Demers et al.
(\cite{demers04}); (3) Mateo (\cite{mateo98}); (4) Jarrett et al.
(\cite{jarrett03}); (5) Sakai et al. (\cite{sakai99}); (6) Vaduvescu
et al. (\cite{vaduvescu07}); (7) Gil de Paz et al. (\cite{gil03});
(8) Huchtmeier et al. (\cite{huchtmeier79}); (9) Wilson et al.
(\cite{wilson96}); (10) Garnett (\cite{garnett02}); (11) Lequeux et
al. (\cite{lequeux79}); (12) Garnett (\cite{garnett90}); (13) Richer
el al. (\cite{richer01}); (14) MG09; (15) Lozinskaya et al.
(\cite{loz09}); (16) Kim et al. (\cite{kim09}); (17) Kennicutt et
al. (\cite{kennicutt94}); (18) Melisse \& Israel (\cite{melisse94});
(19) Bell (\cite{bell03}); (20) White \& Becker (\cite{white92}).}
\end{center}
\end{table}

\subsection{Baryonic components}

The hydrogen gas associated with IC10 extends far beyond its optical
size.  The inner galaxy is located within an extended, complex, and
counter-rotating envelope (Huchtmeier \cite{huchtmeier79}; Shostak
\& Skillman \cite{shostak89}) revealed by radio observations. The
\hi\ distribution has several large holes, possibly produced by SNe
explosions (Wilcots \& Miller \cite{wilcots98}).

Its atomic hydrogen mass has been estimated to be $M_{{\rm
HI}}\sim(1.7-2.3)\times10{^8}~M_{\odot}$ (cf. Huchtmeier
\cite{huchtmeier79}; Huchtmeier \& Ritcher \cite{huchtmeier88};
Wilson et al. \cite{wilson96}; Garnett \cite{garnett02}; Vaduvescu
et al. \cite{vaduvescu07}) and its stellar mass  to be
$M_{star}=(4-6)\times10{^8}~M_{\odot}$ (cf. Sakai et al.
\cite{sakai99}; Jarrett et al. \cite{jarrett03}; Vaduvescu et al.
\cite{vaduvescu07}). The molecular gas constitutes a small fraction
of the total gas, having a mass of $M_{\rm
H_2}\sim1\times10^7~M_{\odot}$ (van den Bergh \cite{vdb00}). Thus,
its total gas fraction is $\sim0.22-0.36$. Finally, its dynamical
mass is $1.7\times10^9~M_{\odot}$ (Mateo \cite{mateo98}). In Table
\ref{Tab:obs}, the main properties of IC10 are summarized.

\subsection{The star formation rate}

Several studies of IC10 have identified stellar populations with
different ages (Massey \& Armandroff \cite{massey95}; Sakai et al.
\cite{sakai99}; Borissova et al. \cite{borissova00}; Sanna et al.
\cite{sanna08}). Deep HST colour-magnitude diagrams have uncovered
both intermediate-age, red clump stars, and old red horizontal
branch (RGB) stars (Sanna et al. \cite{sanna09}). Tracers of the
young stellar populations are \hii~regions (Hodge \& Lee
\cite{hodge90}) and Wolf Rayet (WR) stars (e.g., Crowther et al.
\cite{crowther03}). In particular, WR stars are present in IC10 with
the largest number per unit luminosity in the LG (Massey et al.
\cite{massey92}; Massey \& Armandroff \cite{massey95}). On the other
hand, the old/intermediate-age stellar populations are traced by the
PNe and carbon stars (Magrini et al. \cite{magrini03}; Demers et al.
\cite{demers04}).

IC10 is remarkable because of its present-time high SFR. The current
SFR was estimated from H$\alpha$, far-infrared (FIR), and 1.4~GHz
imaging, while the intermediate age SFR was inferred from PN
counting (MG09). The H$\alpha$ flux implies a SFR of
$\sim0.2~M_{\odot}$yr$^{-1}$ after an average correction for the
internal extinction, but it could be as high as
$0.6~M_{\odot}$yr$^{-1}$ if the largest extinction values available
in the literature are adopted (Leroy et al. \cite{leroy06}). In
addition, the number of WR stars provides an independent estimate of
the present SFR. A comparison between  the number of WRs in IC10 and
a similar galaxy such as SMC-- $\sim$100, among them 30 (12)
confirmed spectroscopically by Crowther et al. (2003) (Massey et al.
\cite{massey03}), indicates that the current SFR of IC10 is $3-4$
times higher than that of SMC, i.e., $\sim0.5~M_{\odot}$yr$^{-1}$
(Leroy et al. \cite{leroy06}). On the other hand, the FIR and radio
continuum observations infer lower estimates of the SFR
($\sim0.05~M_{\odot}$yr$^{-1}$; Thronson et al. \cite{thronson90};
White \& Becker \cite{white92}; Bell \cite{bell03}), probably due to
the escape of UV photons or/and to a low dust content.

The mass of the stellar  population that produced  the observed PNe
was estimated to be $1.2\times10^8~M_{\odot}$ by MG09 from extended
PN counting in the whole optical disk of IC10. From that, they
derived a lower limit to the SFR during the period of time when IC10
PN progenitors were born, i.e. $7-10$ Gyr ago. The SFR value was
estimated  to be $\sim0.02-0.04~M_{\odot}$ yr$^{-1}$.

\subsection{Chemical abundances}

Measuring the elemental abundances of populations with different
ages are necessary to constrain the chemical evolution of a galaxy.
PNe and \hii~regions are useful tools in this sense, having similar
spectra, and thus allowing the determination with common sets of
observations and analysis techniques of the same elemental
abundances.

Basically, they are representative of the ISM composition at two
different epochs in the galaxy lifetime. The progenitor stars of
PNe, the low- and intermediate-mass stars
$1~M_{\odot}<M<8~M_{\odot}$, do not modify the composition of O, Ne,
S, and Ar in the material ejected during and after the asymptotic
giant branch (AGB) phase.  Thus the PN abundances of these elements
are characteristic of the composition of the ISM at the epoch when
the PN progenitor was formed. On the other hand, He, N, and C are
modified in the ejecta, because they are processed during the
lifetime of the progenitor stars. In the case of IC10, the observed
PNe element abundances are consistent with those of an old
population. The low N/O ratios and the undetectable \heii\ emission
lines rule out high mass progenitor stars.  Therefore, their
progenitor stars are of mass $M<1.2~M_{\odot}$, and were therefore
born during the first half of the age of the Universe (see MG09 for
details). At the same time, the chemical abundances of \hii~regions
are representative of the current composition of the ISM. Hence the
abundances of the $\alpha$-elements, such as O, Ne, S, and Ar in
\hii~regions and PNe give us a quantitative measurement of the ISM
chemical enrichment from the epoch of PN progenitor formation to the
present.

Chemical abundances of \hii~regions in IC10 have been obtained by
several authors, starting with Lequeux et al. (\cite{lequeux79}),
then recomputed  by  Skillman et al. (\cite{skillman89}), by Garnett
et al. (\cite{garnett90}) with updated atomic data and ionization
correction factors relative to those used by Lequeux et al.
(\cite{lequeux79}), and by the spectroscopic observations of Richer
et al. (\cite{richer01}). The observations by MG09 enlarged the
number of studied \hii~regions in whose spectra  the electron
temperature can be measured, and complemented them with PN
observations. Spectroscopic observations of the ionized gas were
obtained by Lozinskaya et al. (\cite{loz09}), who derived chemical
abundances only using bright-line ratios.

The chemical abundances used to constrain our chemical evolution
model are those of MG09. We adopted this uniform sample to ensure
consistency between the \hii~region and PN chemical abundance
determinations. Moreover, the chemical abundances of  \hii~regions
observed by MG09 in common with previous surveys agree with their
determinations. The average chemical abundances of PNe and
\hii~regions by MG09  are reported in Table~\ref{Tab:obsabu}. In the
following sections, we adopt oxygen and sulphur abundance in PNe as
tracers of the past ISM composition, while we do not consider helium
and nitrogen since their abundance might be altered by evolutionary
processes occurring in the PN progenitors. We use the chemical
abundances of \hii~regions to constrain our chemical evolution
models at the present time.

\linespread{1.4}
\begin{table*}
\caption{Average chemical abundances of PNe and \hii~regions in IC10 from MG09}
\begin{center}
\begin{tabular}{lllllllllllll}
\hline
\hline
                      &He/H     & O/H($\times10^{-4}$) &12+log(He/H)   & 12+log(O/H)   & 12+log(N/H)   & 12+log(S/H)   & log(N/O)       & log(S/O)\\
\hline
PNe\tablefootmark{a}  &0.107$\pm$0.032 & 2.07$\pm$1.30 &11.01$\pm$0.14 & 8.22$\pm$0.36 & 7.23$\pm$0.30 & 6.29$\pm$0.22 & -1.08$\pm$0.20 &-2.02$\pm$0.40\\
\hii~regions\tablefootmark{a} &0.107$\pm$0.031 & 2.31$\pm$1.59 &11.02$\pm$0.12 & 8.28$\pm$0.29 & 7.41$\pm$0.39 & 6.76$\pm$0.31 & -0.87$\pm$0.51 &-1.52$\pm$0.23\\
\hline
\end{tabular}
\tablefoot{ \tablefoottext{a}{Chemical abundances computed by MG09
also including upper limit electron temperature determinations.}}
\end{center}
\label{Tab:obsabu}
\end{table*}
\linespread{1.1}

\section{The chemical evolution model}

The chemical evolution model adopted for IC10 is the one of Yin et
al. (\cite{yin10}) which is an update of the Bradamante et al.
(\cite{bradamante98}) model for dwarf irregular galaxies. In this
model, we can include different SFHs, either a sequence of short
starbursts followed by quiescent periods or a continuous, but very
mild, star formation. The model takes into account the feedback from
 SNe and stellar winds giving rise to a galactic outflow
when the gas thermal energy exceeds the binding energy of gas. A
dark matter halo is also considered for which the dark matter is 10
times more massive than the luminous one, but roughly 3 times more
extended than the luminous matter (see Yin et al. \cite{yin10} for
details).

The time evolution of the fractional mass of element $i$ in the gas,
$G_i$, is described by the equations
\begin{equation}\label{eq:Gi}
\dot{G}_i=-\psi(t)X_i(t)+R_i(t)+\dot{G}_{i,inf}(t)-\dot{G}_{i,out}(t),
\end{equation}
where $G_i(t)=M_g(t)X_i(t)/M_L(t_G)$ is the gas mass in the form of
an element $i$ normalized to the total baryonic mass $M_L$ at the
present day $t_G=13$~Gyr, $M_g(t)$ is the gas mass at time $t$, and
$X_i(t)$ represents the mass fraction of element $i$ in the gas,
i.e. abundance by mass. If we use the quantity
$G(t)=M_g(t)/M_L(t_G)$ to represent the total fractional mass of
gas, $X_i(t)$ can be expressed by $G_i(t)/G(t)$. The four items on
the right hand side of Eq. (\ref{eq:Gi}) represent the mass change
of the element $i$ caused by the formation of new stars
$\psi(t)X_i(t)$, the material returned by either stellar winds or
SNe explosions $R_i(t)$, the infall of primordial gas
$\dot{G}_{i,inf}(t)$, and the outflow $\dot{G}_{i,out}(t)$,
respectively. The SFR $\psi(t)$ is assumed to be
\begin{equation}\label{eq:psi}
  \psi(t)=\epsilon G(t),
\end{equation}
where $\epsilon$ is the star formation efficiency (SFE) expressed in
unit of Gyr$^{-1}$, and is a free parameters in our work. The
accretion rate of an  element $i$ follows the equation
\begin{equation}\label{eq:inf2}
 \dot{G}_{i,inf}(t)=X_{i,inf}A e^{-t/\tau},
\end{equation}
where $X_{i,inf}$ represents the abundances of the element $i$ in
the infalling gas that we assume to be primordial (without metals),
$A$ is the normalization constant constrained  to reproduce the
present time total mass, and $\tau$ is the accretion timescale. The
rate of gas lost via galactic winds for each element is assumed to
be proportional to the amount of gas present at the time $t$ given
by
\begin{equation}\label{eq:outf}
  \dot{G}_{i,out}(t)=w_i\lambda G(t) X_{i,out}(t),
\end{equation}
where $X_{i,out}(t)$, the abundance of the element $i$ in the wind,
is assumed to be identical to the abundance $X_{i}(t)$ in the ISM,
the parameter $\lambda$ describes the efficiency of the galactic
wind and is expressed in Gyr$^{-1}$ like $\epsilon$, $w_i$ is the
efficiency weight of each element, hence $w_i\lambda$ is the
effective wind efficiency of the element $i$, and $\lambda$ and
$w_i$ are the other two free parameters in our model. In this work,
we have studied two kinds of winds: the normal wind and the
metal-enhanced wind. In the case of the normal wind, all elements
are lost in the same way, i.e., $w_i=1$ for all elements; however,
winds, in which metals are lost preferentially relative to H and He,
are referred to as ``metal- enhanced'' winds. The assumed initial
mass function (IMF) is the Salpeter (\cite{salpeter55}) one, and the
adopted stellar yields are those of Woosley \& Weaver
(\cite{woosley95}) for massive stars and van den Hoek \& Groenewegen
(\cite{van97}) for low- and intermediate-mass stars, both of which
are metallicity-dependent.

\section{Model results}
Since we do not know a priori the SFH of IC10, we computed several
models  by varying the prescriptions for star formation. In
particular, we developed models with either different numbers of
bursts or continuous SF. We refer to the models with several bursts
of duration longer than the interburst phase as models with
``gasping SF'', and models with short bursts (durations shorter than
1 Gyr) as ``bursting'' models. The galaxy is assumed to have been
formed by the continuous infall of primordial gas. In addition, we
tested models with and without galactic wind.

\subsection{Models without galactic winds}
For comparison purposes we developed models without feedback, thus
without galactic winds. In Table~\ref{Tab:nowd}, we show the model
parameters in the star-bursting and gasping cases, both without
galactic winds: the SFE $\epsilon$, the time $t$ corresponding to
the middle of each starburst, the duration $d$ of each burst, and
the timescale for the infall $\tau$ are indicated for each model.

\begin{table}[!t]
\centering \caption{Parameters of models without wind}
\label{Tab:nowd}
\begin{tabular}{c|ccllc}
\hline \hline
Model & $\epsilon$ & n  & $t$\tablefootmark{a}& d                & $\tau$\\
name  & Gyr$^{-1}$ &    & Gyr                 & Gyr              & Gyr \\
\hline
1     & 0.1        & 3  & 2/5/12.5            & 1/1/1            & 0.5 \\
2     & 0.1        & 3  & 2/5/12.5            & 1/1/1            & 8 \\
3     & 0.3        & 10 & 2/3/4/5/6/7/8/      & 0.4/0.1/.../0.1  & 8 \\
     &            &    & 9/10.5/12.95        &                  &   \\
4     & 0.5        & 3  & 2/7/12.5            & 1/7/1            & 8 \\
5     & 0.5        & 10 & 2/3.5/5/6.5/7.5/8.5/& 0.5/0.5/.../0.5  & 8 \\
     &            &    & 9.5/10.5/11.5/12.75 &                  &   \\
\hline \hline
\end{tabular} \\
\tablefoot{\tablefoottext{a}{the middle time of each burst.}}
\end{table}

\begin{figure}[!t]
\centering
\includegraphics[width=0.48\textwidth]{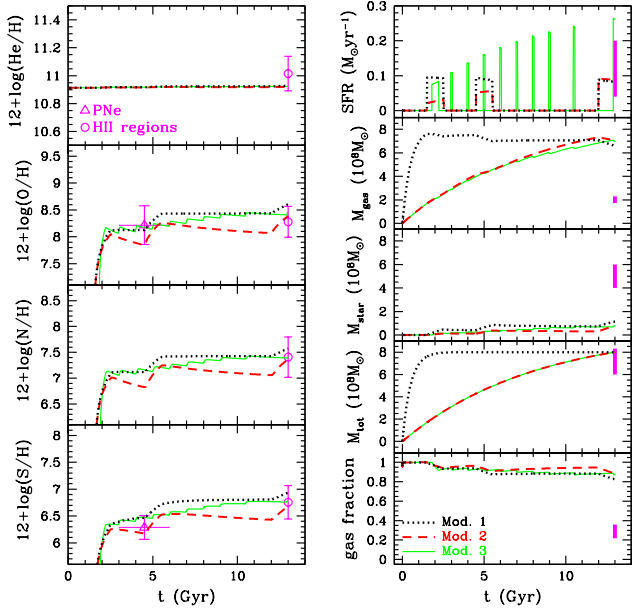} \\
\caption{Time evolutions predicted by the models without winds. The
left column shows the evolution of He, O, N, S abundances, the
magenta points show the average abundances of PNe ({\it open
triangles}, the horizontal lines from 3 to 6 Gyr denote the age
range  where PNe could be) and \hii~regions ({\it open circles}) in
Table \ref{Tab:obsabu}; the right column shows the evolution of the
total SFR, total gas mass, total stellar mass, total baryonic mass,
and gas fraction, the magenta bars at 13 Gyr represent the range of
the observational data (Table \ref{Tab:obs}). {\it Black-dot lines}:
model 1, 3 bursts with short infall timescale ($\tau=0.5$ Gyr); {\it
red-dashed lines}: model 2, 3 bursts with long infall timescale
($\tau=8$ Gyr); {\it green-solid lines}: model 3, 10 bursts with
long infall timescale ($\tau=8$ Gyr).  } \label{Fig:evonowd}
\end{figure}

\begin{figure}[!t]
\centering
\includegraphics[width=0.45\textwidth]{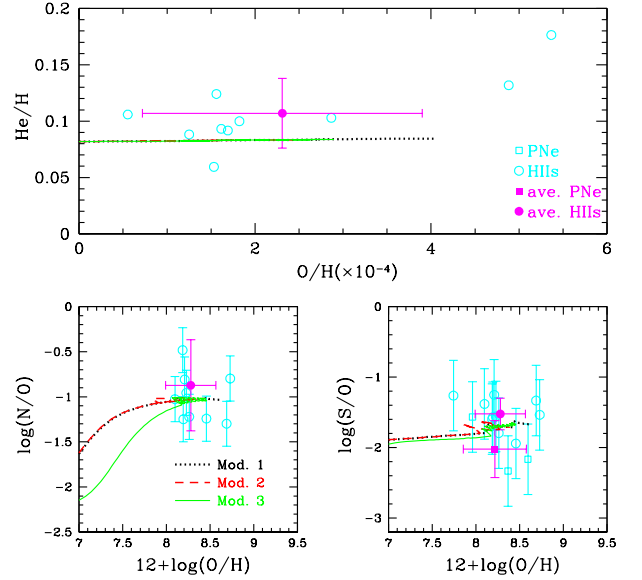} \\
  \caption{Abundance ratios predicted by the models without winds.
The cyan {\it squares} and {\it circles} are the observed abundance
of PNe and \hii~regions (MG09), and the {\it filled magenta points}
are their average chemical abundances (Table \ref{Tab:obsabu}). The
lines are for the same models as in Fig.~\ref{Fig:evonowd}.}
\label{Fig:XOnowd}
\end{figure}

Models 1 and 2, assuming that IC10 has suffered 3 bursts (2 bursts
during the first half of the age of the Universe, 2 and 5 Gyr, and a
third occurring roughly at the present time, namely at  12.5 Gyr)
predict acceptable present-day abundances (the left column of
Fig.~\ref{Fig:evonowd}), if one adopts  a low SFE ($\epsilon=0.1$)
and a reasonably long burst duration ($d=1$ Gyr). For a short infall
timescale ($\tau=0.5$ Gyr, model 1), the abundances of different
elements increase during the burst and maintain that level in the
interburst phase, whereas for a long infall timescale ($\tau=8$ Gyr,
model 2) the abundances decrease during the interburst phase owing
to the dilution of the infalling gas, thus the present-time
abundances are lower than those predicted by the fast accretion
model, even if all the other parameters are the same. Since there is
no SF in these models (models 1 and 2) between 5.5 and 12.0 Gyr and
we do not predict the existence of stars with ages between 7.5 and 1
Gyr, at variance with observations, we compiled another model with
10 short bursts ($d\sim0.1$ Gyr) and a long infall timescale
($\tau=8$ Gyr, to avoid a too high metallicity at the present time).
This model (model 3) also shows acceptable abundance evolutionary
tracks.

However, all of these models form too few stars and a lot of gas
remains, hence the predicted gas fraction ($0.8 - 0.9$) at the
present time is very high and inconsistent with the observations
($0.22 - 0.36$), as shown in the right column of
Fig.~\ref{Fig:evonowd}. The abundance ratios (e.g., N/O, S/O)
predicted by these models are consistent with the average trend of
observations, but cannot explain the scatter in the data (see
Fig.~\ref{Fig:XOnowd}). However, the spread in the oxygen abundance
of \hii~regions seems to be real, and was confirmed not only by the
 study of MG09, but also by earlier observations (Lequeux et
al. 1979, Hodge \& Lee \cite{hodge90}, Richer et al.
\cite{richer01}). The scale-length of metallicity inhomogeneities is
of $0.2-0.3$ kpc, some clumps exhibit the highest measured
metallicities, and an area with a more constant metallicity, around
12+$\log$(O/H) = 8.2, being located between 0.2 to 0.5 kpc from the
center (e.g. MG09).

The origin of the inhomogeneities depends on the efficiency of the
mixing mechanisms in dIrrs. During the quiescent phases after the
star formation episodes, the chemical elements ejected during the
final stellar evolution phases cool until reaching the temperature
of the ISM, and then mix with the rest of the ISM. The mixing is
supposed to be homogeneous and efficient throughout the galaxy, and
produced by epicyclic or radial mixing, or superbubble expansion
(Roy \& Kunth \cite{roy95}). However, the mixing mechanisms
operating on large scales are  less efficient for dIrrs because of
the longer timescales of the gas mixing, and thus inhomogeneities
can survive in some cases. In addition to IC10, \hii~regions with
different chemical compositions were observed in the dIrr galaxies
WLM (Hodge \& Miller 1995), NGC5253 (Kobulnicky et al. 1997), and IC
4662 (Hidalgo-G\'amez et al. 2001). Weshow in Sect. \ref{sec_wind}
how models that take account of chemically enhanced winds are able
to explain the observed metallicity scatter.

\begin{figure}[!t]
\centering
\includegraphics[width=0.48\textwidth]{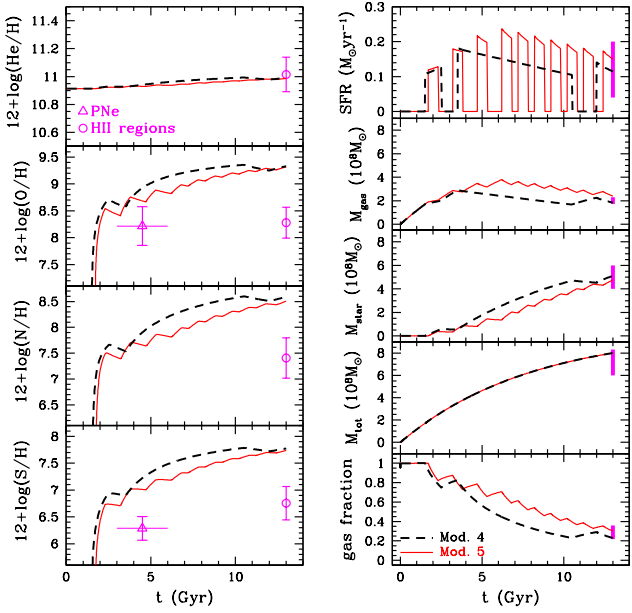} \\
  \caption{Time evolutions predicted by the models 4 and 5 without winds.
The left column shows the evolution of He, O, N, S abundances, and
the right column shows the evolution of the total SFR, total gas
mass, total stellar mass, total baryonic mass, and gas fraction. The
observational constraints are the same as in Fig.~\ref{Fig:evonowd}.
{\it Black-dash lines}: model 4, 3 bursts with higher SFE
($\epsilon=0.5$); and {\it red-solid lines}: model 5, 10 bursts with
higher SFE ($\epsilon=0.5$). Both models adopt a long infall
timescale ($\tau=8$ Gyr).  } \label{Fig:evonowd2}
\end{figure}

\begin{figure}[!t]
\centering
\includegraphics[width=0.45\textwidth]{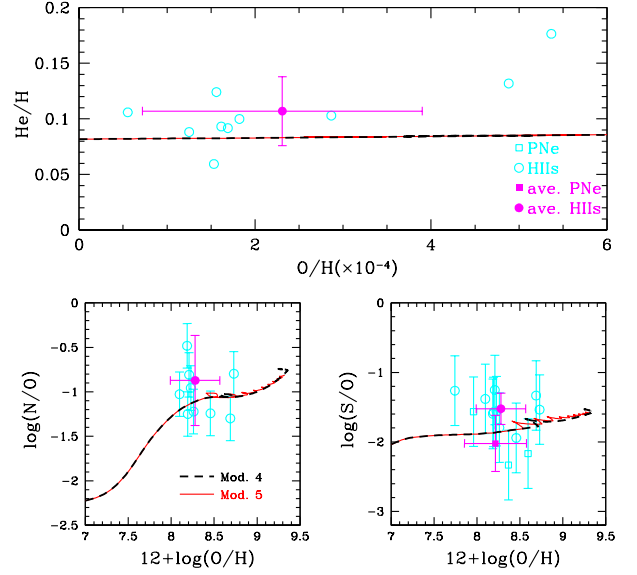} \\
  \caption{Abundance ratios predicted by the models 4 and 5 without winds.
The points are the observed abundance of PNe and \hii~regions
(MG09), same as in Fig.~\ref{Fig:XOnowd}. The lines are for the same
models as in Fig.~\ref{Fig:evonowd2}.} \label{Fig:XOnowd2}
\end{figure}

If we increase the SFE to consume more gas ($\epsilon=0.5$, models 4
and 5), the system reaches very high abundances, even if the long
infall timescale is adopted. In Figs.~\ref{Fig:evonowd2} and
~\ref{Fig:XOnowd2}, we show both the ``gasping'' and ``bursting''
model results. Both of them can reproduce the observed gas mass and
gas fraction, but fail to reproduce most of the abundances.

\subsection{Models with winds}
When stellar feedback is included, galactic winds naturally develop
in small galaxies. Here we describe the results of models with
winds. As we have previously shown, if SFE is low the system is too
gas-rich, whereas if SFE is high the system is too metal-rich.
Therefore, the galactic wind should play an important role in the
evolutionary history of IC10.

The rate of mass loss for the element $i$ is $w_i \lambda G_i$,
where $G_i(t)$ is the gas mass in the form of the element $i$
normalized to the total baryonic mass at the present day, $\lambda$
is the wind efficiency and is the same for all the elements, and
$w_i$ is the weight of the element $i$. In Table~\ref{Tab:wd}, we
show the parameters adopted for the wind models (both normal and
metal-enhanced). In particular, $w1$ and $w2$ are the models with
normal winds, whereas all models labelled $mw$ have metal-enhanced
winds.

\subsubsection{Normal winds}

\begin{figure}[!t]
\centering
\includegraphics[width=0.48\textwidth]{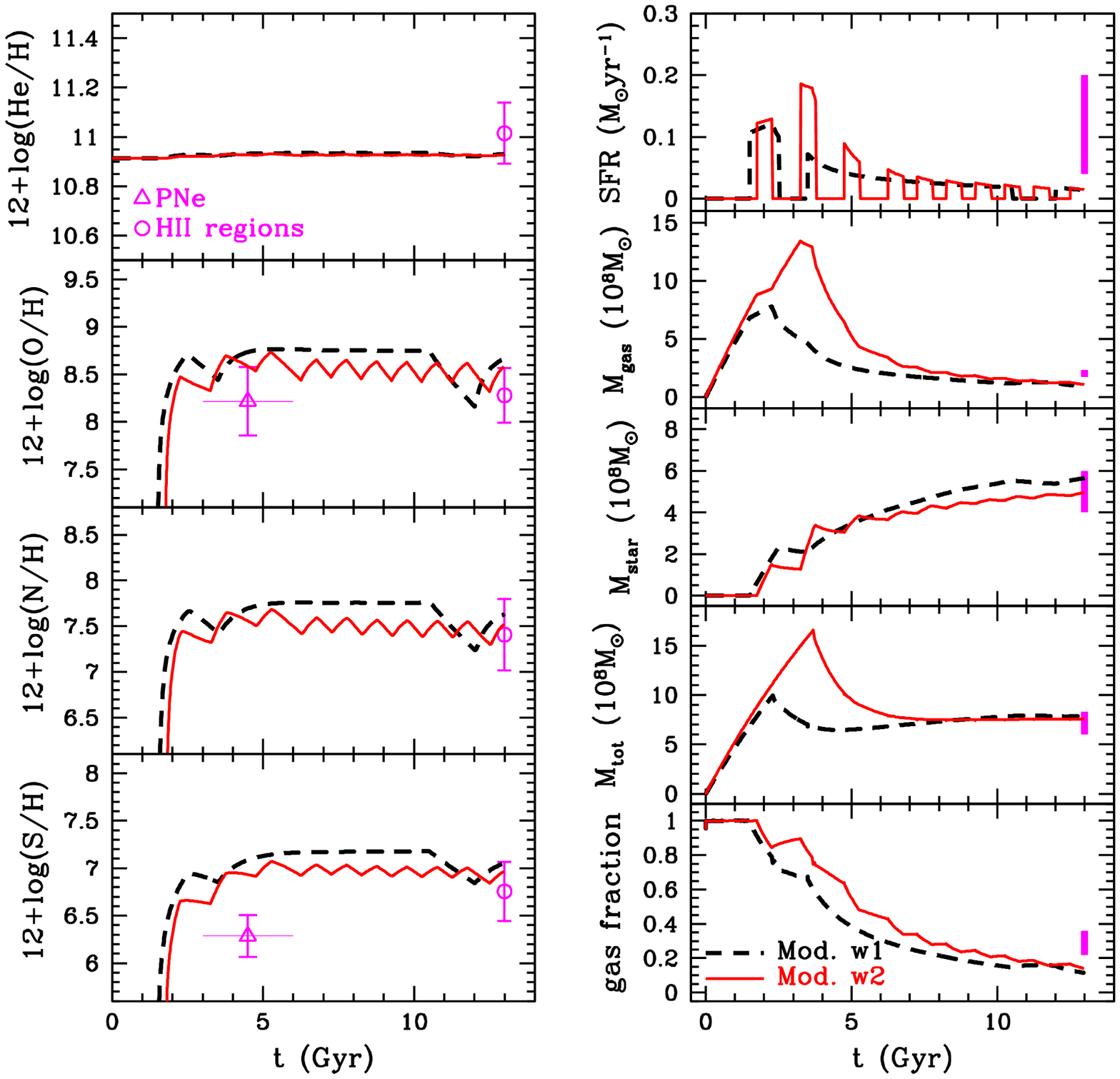} \\
  \caption{Time evolutions predicted by the models with normal winds.
The left column shows the evolution of He, O, N, S abundances, and
the right column shows the evolution of the total SFR, total gas
mass, total stellar mass, total baryonic mass, and gas fraction. The
observational constraints are the same as in Fig.~\ref{Fig:evonowd}.
{\it Black-dash lines}: model w1, 3 bursts ($\epsilon=0.5,
M_{inf}=32\times10^8$~\ms); and {\it red-solid lines}: model w2, 10
bursts($\epsilon=0.5, M_{inf}=36\times10^8$~\ms). Both models adopt
long infall timescale ($\tau=8$ Gyr).  } \label{Fig:evonormwd}
\end{figure}

\begin{figure}[!t]
\centering
\includegraphics[width=0.45\textwidth]{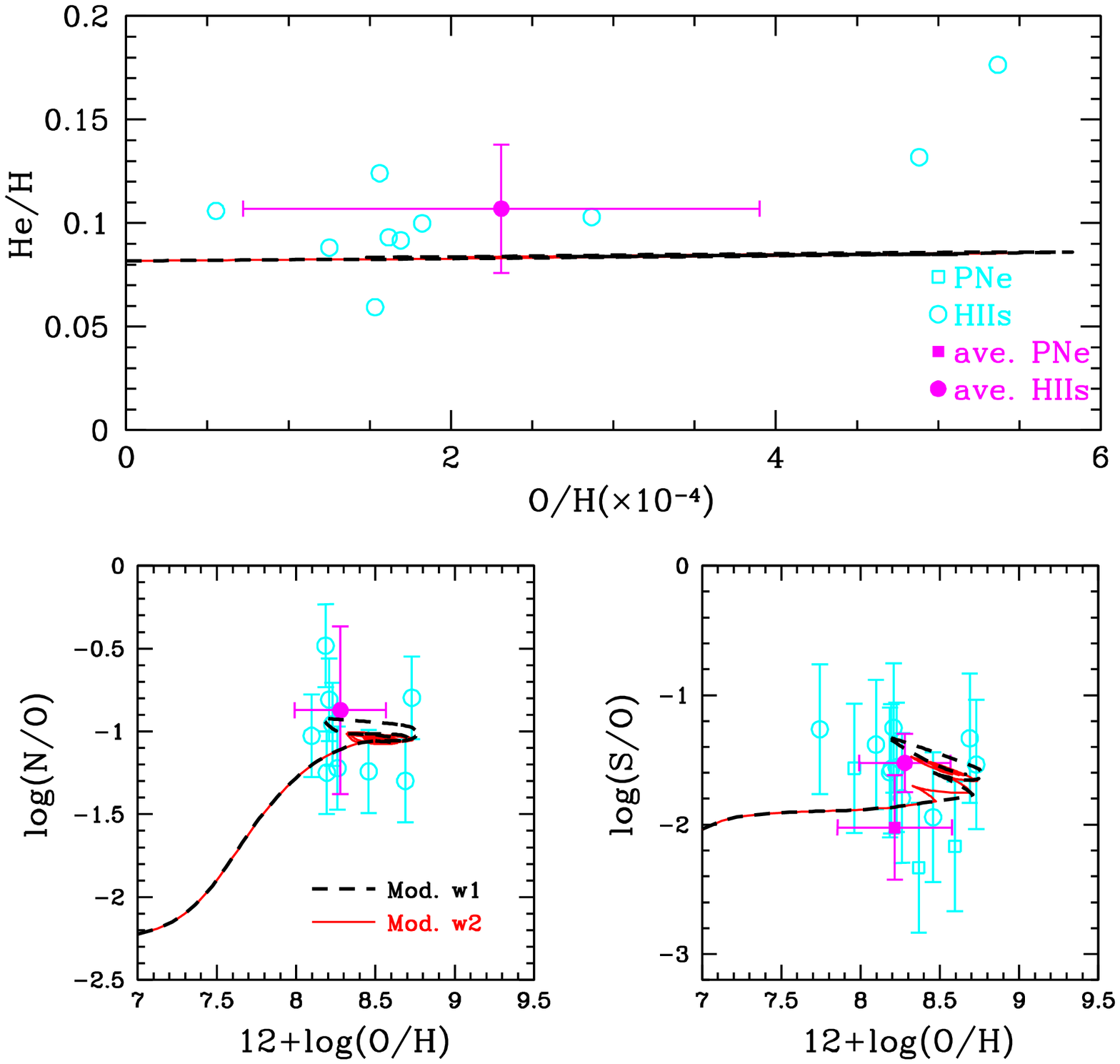} \\
  \caption{Abundance ratios predicted by the models with normal winds.
The points are the observed abundance of PNe and \hii~regions
(MG09), same as in Fig.~\ref{Fig:XOnowd}. The lines are for the same
models as in Fig.~\ref{Fig:evonormwd}.} \label{Fig:XOnormwd}
\end{figure}

In Figs.~\ref{Fig:evonormwd} and ~\ref{Fig:XOnormwd}, we show the
results for normal winds, i.e.,  when all the elements are lost at
the same rate, as opposed to metal-enhanced winds, i.e. when metals
are preferentially lost relative to H and He. In other words, in
normal winds we assume that $w_i$ is the same for all the elements.

In this situation, since the wind  blows away a large amount of gas,
the total mass of infalling gas must increase to reproduce the
observed total baryonic mass inside the galaxy at the present day.

\begin{table*}[!t]
\centering \caption{Parameters of models with wind ($\tau=8$ Gyr)}
\label{Tab:wd}
\begin{tabular}{l|ccllcc}
\hline \hline
Model & $\epsilon$ & n  & $t$\tablefootmark{a}& d                & $\lambda$    & $M_{inf}$      \\
name  & Gyr$^{-1}$ &    & Gyr                 & Gyr              & Gyr$^{-1}$ & $10^8$\ms            \\
\hline
w1    & 0.5        & 3  & 2/7/12.5            & 1/7/1            & 1          & 32  \\
w2    & 0.5        & 10 & 2/3.5/5/6.5/7.5/8.5/9.5/10.5/11.5/12.75& 0.5/0.5/.../0.5 & 1 & 36  \\
\hline
mw1   & 0.5        & 3  & 2/7/12.5            & 1/7/1            & 1          & 11  \\
mw2   & 0.5        & 3  & 2/7/12.5            & 1/7/1            & 2          & 13  \\
mw3   & 0.2        & 1  & 6.6                 & 12.8             & 1.5        & 15   \\
mw4   & 0.3        & 1  & 8.5                 & 9                & 1.5        & 13   \\
mw5   & 0.4        & 1  & 10                  & 6               & 1.5        & 12   \\
mw6   & 0.4        & 2  & 5/10                & 0.6/6            & 1.5        & 13   \\
mw7   & 0.4        & 3  & 4.5/9/12.5          & 3/3/1            & 1.5        & 13   \\
mw8   & 0.4        & 4  & 4/7/10/12.5         & 2/2/2/1          & 1.5        & 13   \\
mw9   & 0.4        & 6  & 3/4.5/6/8/10/12.5   & 1/1/1/1.5/1.5/1  & 1.5        & 13   \\
mw10  & 1          & 4  & 3/5/7/12.7          & 0.8/0.8/0.8/0.6  & 1.5        & 13   \\
mw11  & 1          & 8  & 3/4/5/6/7/9/11/12.8 & 0.4/0.4/.../0.4 & 1.5 & 13   \\
mw12  & 1          & 11 & 3/4/5/6/7/8/9/10/11/12/12.9 & 0.2/0.2/0.2/0.2/0.2/0.3/0.3/0.3/0.3/0.3/0.2 & 1.5 & 13   \\
\hline \hline
\end{tabular} \\
\tablefoot{\tablefoottext{a}{the middle time of each burst.}}
\end{table*}

The wind efficiency is assumed to be $\lambda=1$, and both
``bursting'' and ``gasping'' models are tested (see
Figs.~\ref{Fig:evonormwd} and ~\ref{Fig:XOnormwd}). According to our
models, to have a final mass of $8\times10^8$\ms\ we should accrete
 more than $3\times10^9$\ms, which is 4 times higher than the
present day total galactic mass. This implies that in our models an
unrealistically large amount of gas is lost.

In these models, the gas infall rates are also very high, and the
galaxy accumulates its mass rapidly, as we can see from the
evolutionary tracks of gas mass and total baryonic mass. A large
number of stars (about one third of the present stellar mass) have
formed during the first star formation episode resulting in a rapid
increase in metallicity at early galactic epochs. Although the
enrichment is decelerated by the development of the wind, the
elemental abundances are much higher than those observed in PNe. The
present gas mass and gas fraction are lower than those suggested by
observations because of the efficient wind. As a consequence, the
star formation activity declines rapidly and reaches a very low
level at the present day, in contrast to the observed value.

To summarize, the development of a normal wind implies that the
galaxy should have lost an unrealistically large part of its
accreted mass and been enriched very rapidly at early epochs.
Therefore, it is reasonable to assume that mainly metals have been
lost.

\subsubsection{Metal-enhanced winds}
\label{sec_wind}

\begin{figure}[!t]
\centering
\includegraphics[width=0.48\textwidth]{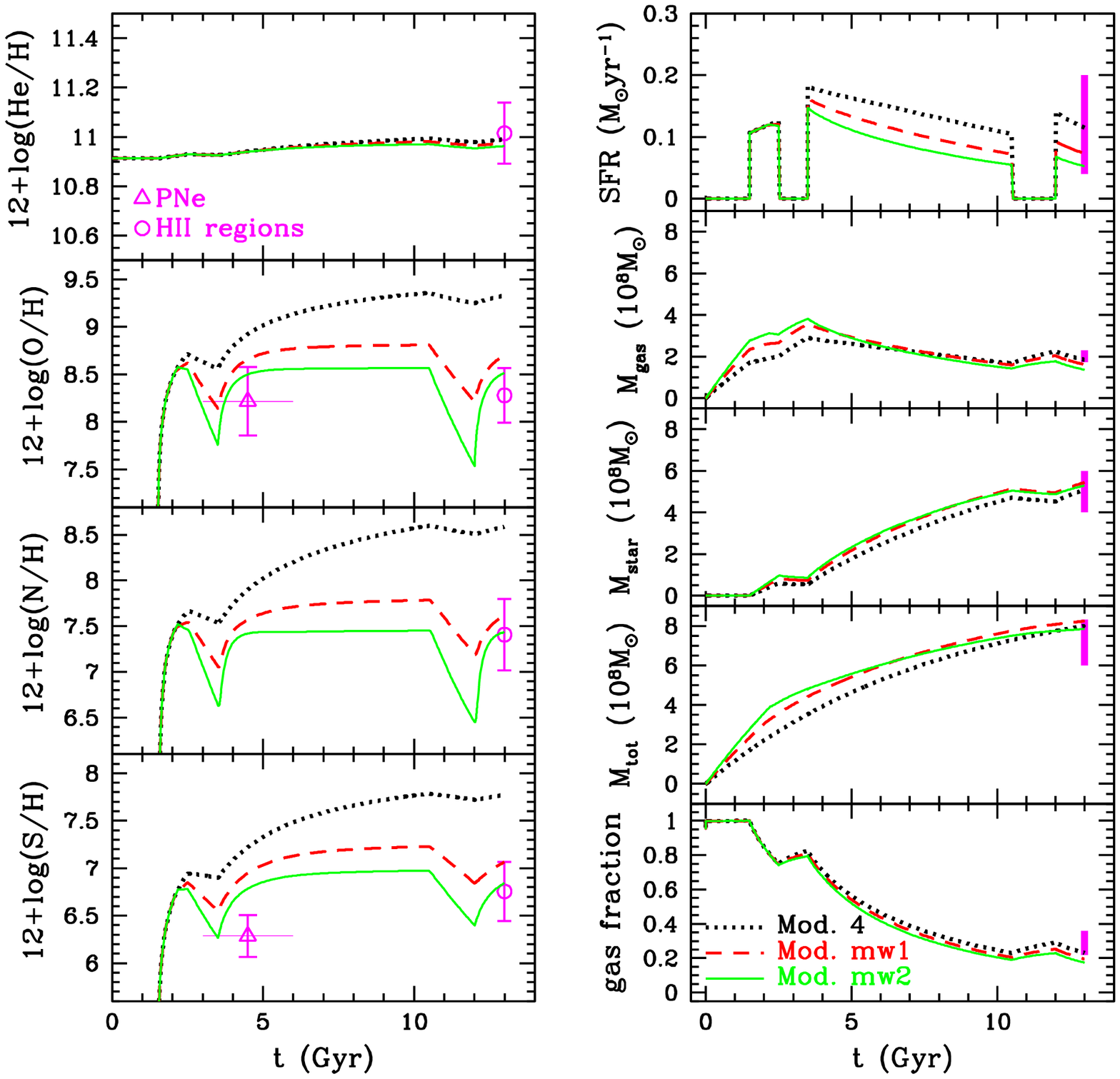} \\
  \caption{Time evolutions predicted by the models with metal-enhanced winds.
The left column shows the evolution of He, O, N, S abundances, and
the right column shows the evolution of the total SFR, total gas
mass, total stellar mass, total baryonic mass, and gas fraction. The
observational constraints are the same as in Fig.~\ref{Fig:evonowd}.
{\it Black-dot lines}: model mw1, $\lambda=0,
M_{inf}=8\times10^8$\ms; {\it red-dash lines}: model mw2,
$\lambda=1, M_{inf}=11\times10^8$\ms; and {\it green-solid lines}:
model mw3, $\lambda=2, M_{inf}=13\times10^8$\ms.}
\label{Fig:evodiffwi}
\end{figure}

\begin{figure}[!t]
\centering
\includegraphics[width=0.45\textwidth]{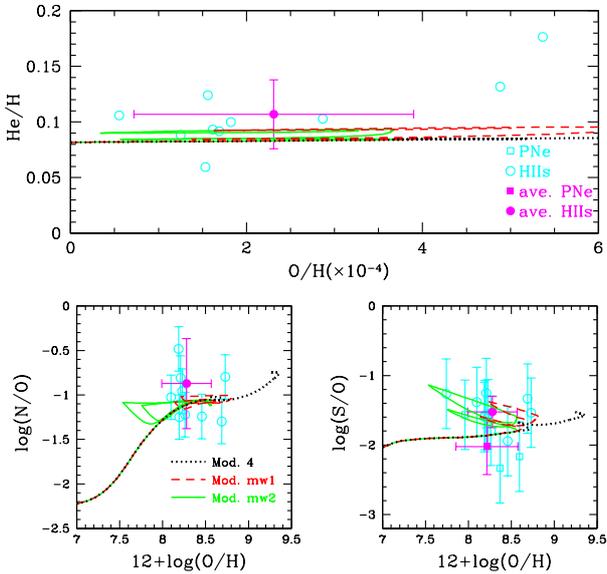} \\
  \caption{Abundance ratios predicted by the models with metal-enhanced winds.
The points are the observed abundance of PNe and \hii~regions
(MG09), same as in Fig.~\ref{Fig:XOnowd}. The lines are for the same
models as in Fig.~\ref{Fig:evodiffwi}.} \label{Fig:XOdiffwi}
\end{figure}

``Metal-enhanced'' winds have been suggested by several authors (Mac
Low \& Ferrara \cite{mac99}; D'Ercole \& Brighenti \cite{dercole99};
Recchi et al. \cite{recchi01}; Fujita et al. \cite{fujita03}; Recchi
et al. \cite{recchi08}). Here, we assume that $w_{\rm H,He}=0.1,
w_{\rm O,S}=0.85, w_{\rm N}=0.97$ (for this particular choice, we
followed Romano et al. \cite{romano06}), which means that H and He
are lost at a very low rate.

In Figs.~\ref{Fig:evodiffwi} and \ref{Fig:XOdiffwi}, we show the
results of different wind efficiencies $\lambda=0, 1, 2$. In
Fig.~\ref{Fig:evodiffwi}, it is clear that the stronger the wind,
the lower the metallicity, and the deeper the minima in the
abundance tracks occurring in interburst time. On the other hand,
the loops in the tracks in Fig.~\ref{Fig:XOdiffwi} are caused by the
galactic wind and can explain the scatter in S/O data. However, the
models are less successful in reproducing the scatter in N/O
abundance ratio, especially for 12+log(O/H)$>8.2$.

Since the observed present-time oxygen abundance, 12+$\log$(O/H), is
mainly between 7.9 and 8.7, implying that the wind efficiency cannot
be too high, we adopted $\lambda=1.5 $ in this paper. This adopted
value of wind efficiency implies a reasonable amount of infalling
gas ($M_{inf}=13\times10^8$ \ms), and also provides a good fit to
the observed masses, as we can see in the next section.

\subsection{The effects of different SF regimes in the metal-enhanced-wind case}

In this section, we describe in more detail the models with
metal-enhanced wind that we developed. We tested three types of SFH:
continuous SF, gasping SF, and bursting SF.

\subsubsection{Continuous star formation}

\begin{figure}[!t]
\centering
\includegraphics[width=0.48\textwidth]{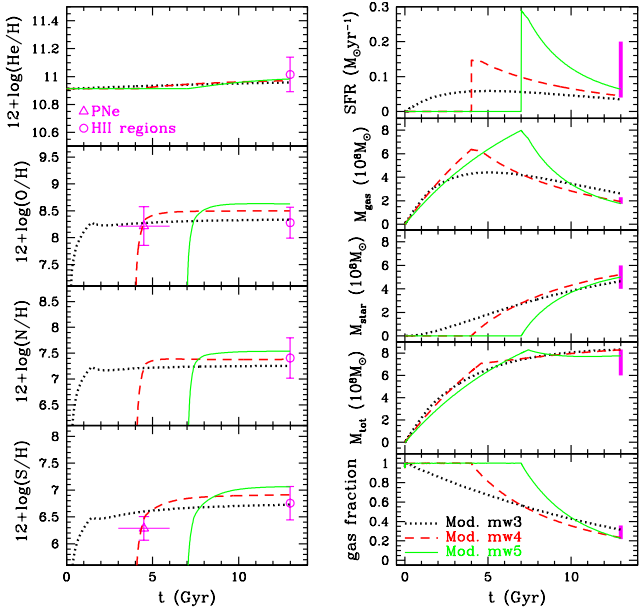} \\
\caption{Time evolutions predicted by the models with continuous
star formation and metal-enhanced winds. The left column shows the
evolution of He, O, N, S abundances, and the right column shows the
evolution of the total SFR, total gas mass, total stellar mass,
total baryonic mass, and gas fraction. The observational constraints
are the same as in Fig.~\ref{Fig:evonowd}. {\it Black-dot lines}:
model mw3, $\epsilon=0.2, d=12.8$ Gyr; {\it red-dash lines}: model
mw4, $\epsilon=0.3, d=9$ Gyr; and {\it green-solid lines}: model
mw5, $\epsilon=0.4, d=6$ Gyr.} \label{Fig:evocontiSF}
\end{figure}

\begin{figure}[!t]
\centering
\includegraphics[width=0.45\textwidth]{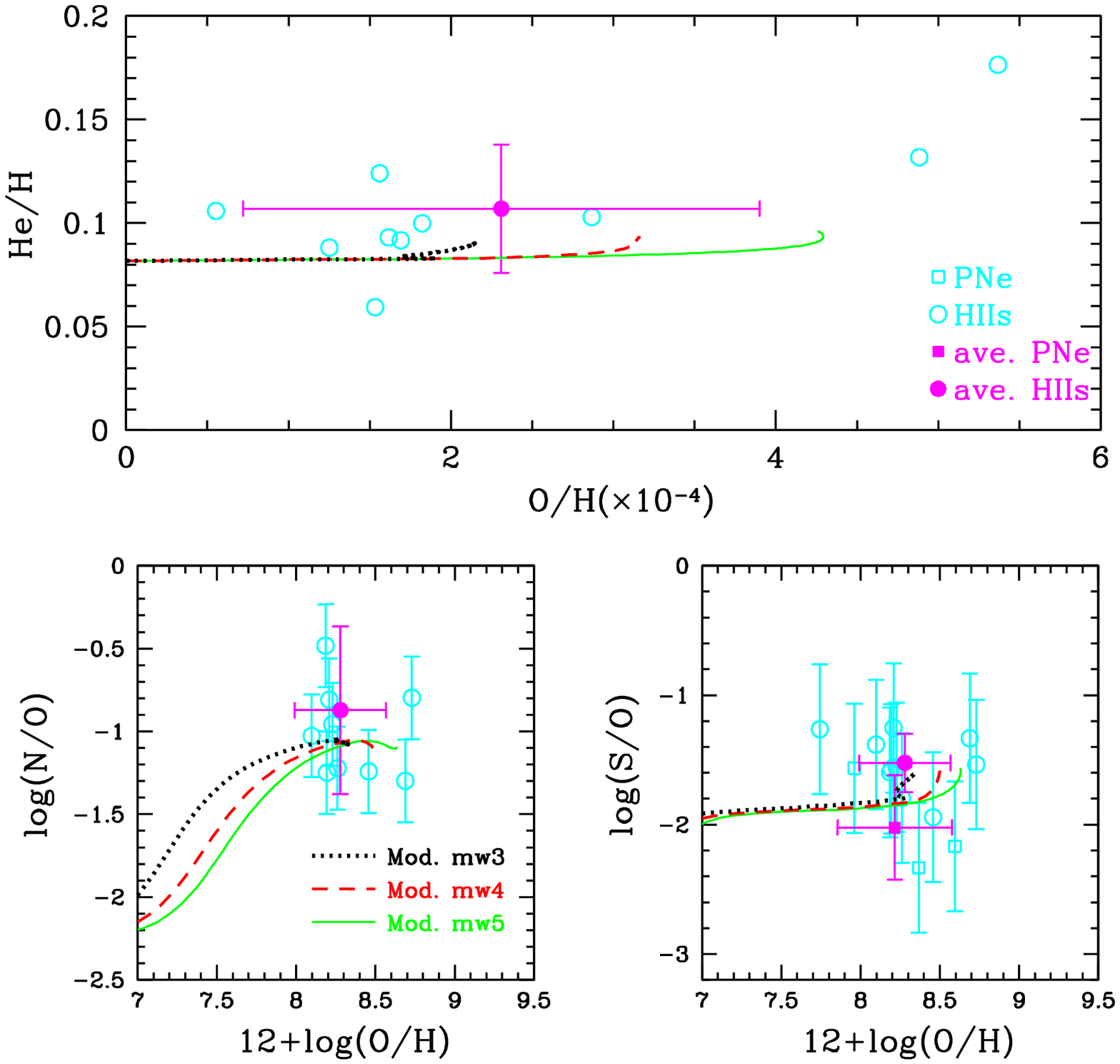} \\
  \caption{Abundance ratios predicted by the models with continuous
star formation and metal-enhanced winds. The points are the observed
abundance of PNe and \hii~regions (MG09), and are the same as in
Fig.~\ref{Fig:XOnowd}. The lines are for the same models as in
Fig.~\ref{Fig:evocontiSF}.} \label{Fig:XOcontiSF}
\end{figure}

We developed and tested some models with continuous star formation
lasting until the present time. Different SFEs were adopted. We
found that if the SFE is low, the star formation duration should be
long enough to form the observed number of stars (12.8 Gyr if
$\epsilon=0.2$, in Figs.~\ref{Fig:evocontiSF} and
~\ref{Fig:XOcontiSF}). On the other hand, if the SFE is higher, one
should assume that the SF starts later and lasts for a shorter time
(6 Gyr if $\epsilon=0.4$,  in Figs.~\ref{Fig:evocontiSF} and
~\ref{Fig:XOcontiSF}). In the former case, where SF is active since
the very beginning of galaxy formation by infall, the wind begins to
remove gas at earlier times (hence a higher total infall mass is
required), and the metallicity increases very slowly, reaching an
acceptable but low present-time value. In the latter case, since we
need to reproduce the present time SFR, the star formation must have
initiated $\sim$6 Gyr ago, and be very high when it begins, since
the galaxy has already accreted a lot of gas. However, in this case
no stars form from 7 to 10 Gyrs ago, which is inconsistent with the
observations. In Fig.~\ref{Fig:XOcontiSF}, we can see that the
continuous star-formation scenario cannot explain the scatter in the
metallicity data.

\subsubsection{Gasping star formation}

\begin{figure}[!t]
\centering
\includegraphics[width=0.48\textwidth]{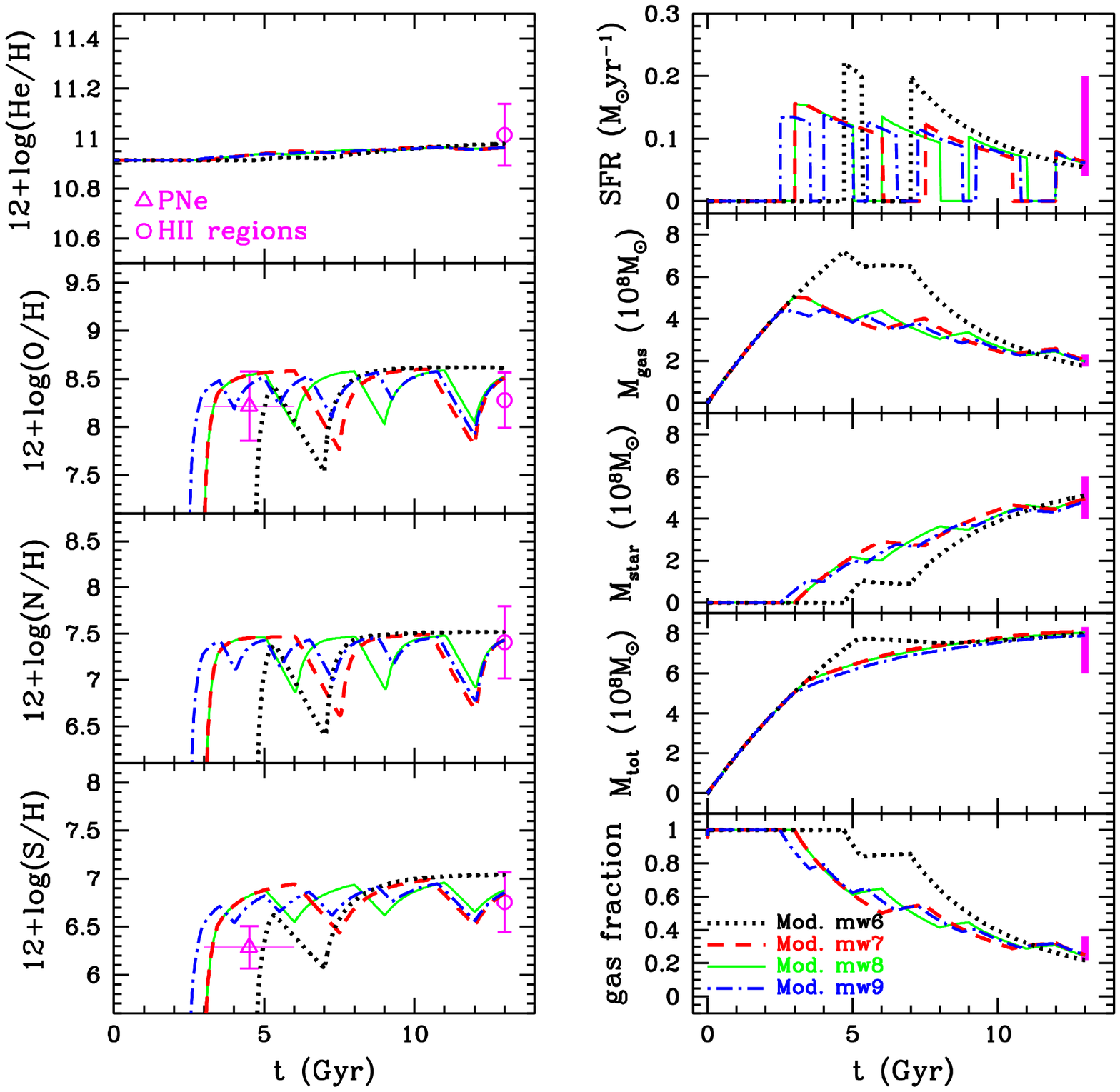} \\
  \caption{Time evolutions predicted by the models with gasping
star formation and metal-enhanced winds. The left column shows the
evolution of He, O, N, S abundances, and the right column shows the
evolution of the total SFR, total gas mass, total stellar mass,
total baryonic mass, and gas fraction. The observational constraints
are the same as in Fig.~\ref{Fig:evonowd}. {\it Black-dot lines}:
model mw6, $n=2$; {\it red-dash lines}: model mw7, $n=3$; {\it
green-solid lines}: model mw8, $n=4$; and {\it blue-dash-dot lines}:
model mw9, $n=6$} \label{Fig:evogaspSF}
\end{figure}

\begin{figure}[!t]
\centering
\includegraphics[width=0.45\textwidth]{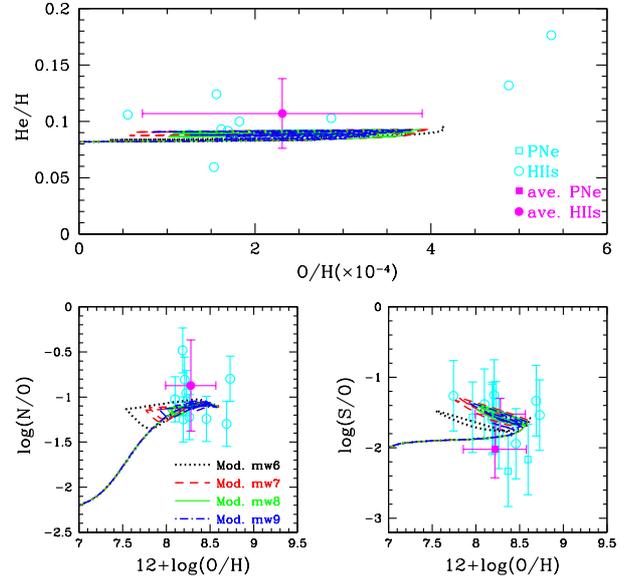} \\
  \caption{Abundance ratios predicted by the models with
gasping star formation and metal-enhanced winds. The points are the
observed abundance of PNe and \hii~regions (MG09), and are the same
as in Fig.~\ref{Fig:XOnowd}. The lines are for the same models as in
Fig.~\ref{Fig:evogaspSF}.} \label{Fig:XOgaspSF}
\end{figure}

When we use the observed present SFR and gas fraction to constrain
the SFE, we find that $\epsilon\sim0.3-0.4$, if SFRs derived from
FIR and 1.4 GHz measurement are used. Here we adopt $\epsilon=0.4,
M_{inf}=13\times10^8$~\ms.

In Figs.~\ref{Fig:evogaspSF} and \ref{Fig:XOgaspSF}, we present our
results for models with different numbers of bursts ($n=2$, 3, 4, 6,
where the last burst is ongoing at the present day). The total
duration of SF is $\sim$7 Gyr in all these 4 cases.

For model mw6 ($n=2$), the first short burst happens at 5 Gyr, then
star formation halts for 2 Gyr, and restarts at $t=7$ Gyr. The first
burst cannot happen too early, otherwise the minimum in metallicity
(mainly caused by wind) will be too deep. In this model, we produce
very few old metal-poor stars, but a bunch of metal-rich and younger
stars, which does not reflect the star formation history of IC10.

In model mw7 ($n=3$), there are three bursts, one during the first
half of the galactic age, one during the second half of the galactic
age, and the last one which remains active at the present time; the
interburst time is about 1.5 Gyr. The results are able to more
closely reproduce the data.

For models mw8 ($n=4$) or mw9 ($n=6$), since the wind and the
infalling gas decrease the metallicity of ISM during the interburst
phase, the metal-poor stars can be formed if the galaxy experiences
more star formation bursts. In these two cases, the models have more
metal-poor stars formed in the first half of the galactic age and
can also fit other observational data. Therefore, they are our
best-fit models.

In Fig.~\ref{Fig:XOgaspSF}, we can see that the gasping
star-formation scenario can reproduce the scatter in the abundance
ratios, especially in the S/O ratio, as we mentioned in Sect. 4.2.2.

\subsubsection{ Bursting star formation}

\begin{figure}[!t]
\centering
\includegraphics[width=0.48\textwidth]{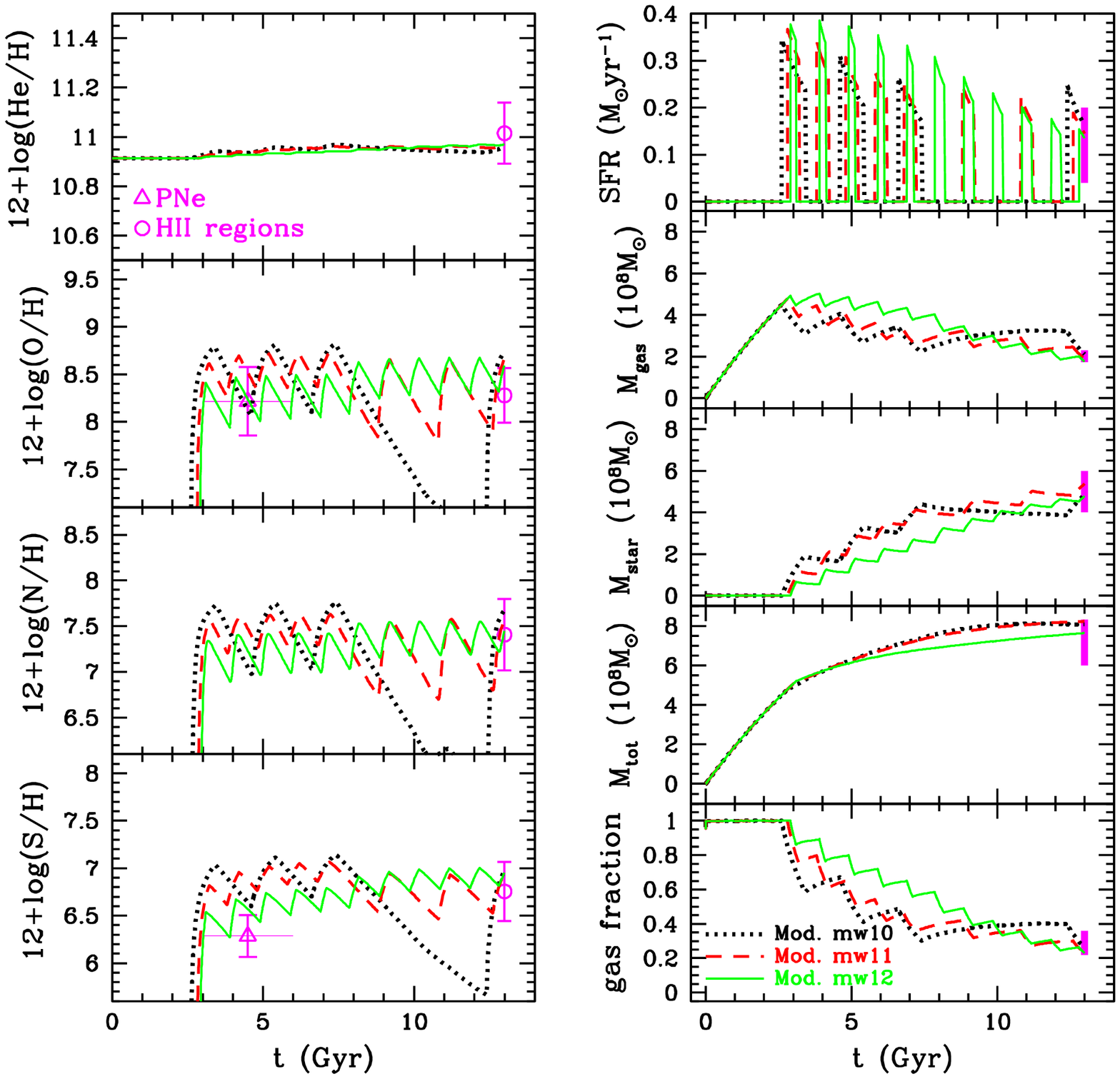} \\
  \caption{Time evolutions predicted by the models with bursting
star formation and metal-enhanced winds.The left column shows the
evolution of He, O, N, S abundances, and the right column shows the
evolution of the total SFR, total gas mass, total stellar mass,
total baryonic mass, and gas fraction. The observational constraints
are the same as in Fig.~\ref{Fig:evonowd}. {\it Black-dot lines}:
model mw10, $n=4$; {\it red-dash lines}: model mw11, $n=8$; and {\it
green-solid lines}: model mw12, $n=11$.} \label{Fig:evoburstSF}
\end{figure}

\begin{figure}[!t]
\centering
\includegraphics[width=0.45\textwidth]{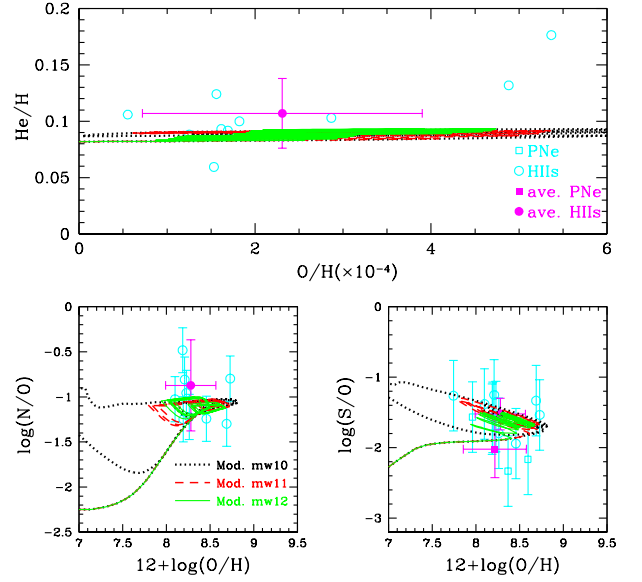} \\
  \caption{Abundance ratios predicted by the models models with
bursting star formation and metal-enhanced winds. The points are the
observed abundance of PNe and \hii~regions (MG09), and are the same
as in Fig.~\ref{Fig:XOnowd}. The lines are for the same models as in
Fig.~\ref{Fig:evoburstSF}.} \label{Fig:XOburstSF}
\end{figure}

In the gasping SF scenario, the model predicts a lot of stars whose
abundances are close to the upper limit of observed PNe. Here a
question arises: is it possible that IC10 is formed by short SF
bursts ($d<1$~Gyr)? If this were true, the SFE should be higher than
in the gasping SF scenario to form enough stars during these short
bursts. We assume here that the SFE $\epsilon=1$, and that the total
SF duration is around 3 Gyr in this case.

In model mw10 ($n=4$), three bursts occur in the first half of the
galactic lifetime when the progenitors of PNe were born, and one
occurs at the present time. A duration $d\sim0.8$ Gyr for each burst
is required to form enough stars. This leads to a relatively high
metallicity after the first 3 bursts, followed by a sharp decrease
in the metallicity during  the long interburst period before the
last burst. In this way, extremely metal-poor stars would have
recently formed, at variance with observations of \hii~regions, as
shown in Fig.~\ref{Fig:XOburstSF}. To avoid these problems, we need
more and shorter bursts.

In model mw11 ($n=8$), there are five bursts before $t=7$ Gyr and
three bursts after. Since the interburst time is short between the
first five bursts, the abundances do not decrease too much. Lots of
stars form between $3-7$ Gyr and basically within the scatter of the
data.

In model mw12 ($n=11$), we assume shorter bursts ($d\sim0.2-0.3$).
Stars formed between $3-6$ Gyr have metallicity consistent with the
observed PNe, but in this model we still have lots of stars formed
at a late age. The results of this model are consistent with
observations.

Compared with the former two SF regimes, the bursting one reproduces
more closely the scatter in the abundance ratios, especially for the
N/O ratio when 10+log(O/H)$>$8.2, as shown in
Fig~\ref{Fig:XOburstSF}. Therefore, model mw12 can also be
considered to be among the best ones.

All these models have SFRs ($\sim0.15-0.2$ \msyr) consistent with
that derived from H$\alpha$ observations, but slightly higher than
the SFR obtained from FIR and 1.4GHz measurements $\sim 0.05-0.07$
\msyr (see Table \ref{Tab:obs}). If we were to accept the SFR value
derived from H$\alpha$, then the bursting SF history might also be
acceptable for IC10. On the other hand, the SFR values obtained from
FIR and 1.4 GHz measurements imply that a gasping SF history is more
likely.

\section{Conclusions}
We have studied the chemical evolution of the dwarf irregular galaxy
IC10 for which detailed abundance data are now available (MG09). We
have adopted a chemical evolution model for dwarf irregulars
developed by Yin et al. (\cite{yin10}). This model includes detailed
metallicity-dependent stellar yields for both massive and low and
intermediate mass stars. The model also includes feedback from SNe
and stellar winds and follows the development of galactic winds. We
have explored three cases: {\em i}) no winds (no feedback); {\em
ii}) normal winds, namely all the gas is lost at the same rate; and
{\em iii}) metal-enhanced winds where the metals are lost
preferentially relative to H and He. We also explored several
regimes of star formation: a) bursting mode with short and long
bursts (a regime we called ``gasping SF''); and b) a continuous but
low SFR.

We computed the evolutions of the abundances of several elements
(He, N, O, S). The parameters of our model were: 1) the number of
bursts; 2) the duration of bursts; 3) the efficiency of star
formation; and 4) the efficiency of the galactic wind rate, assumed
to be proportional to the amount of gas present at the time of the
wind, which can vary from element to element in the case of the
metal-enhanced wind.

The observational constraints were represented by the abundances and
abundance ratios of the above-mentioned elements, by the
present-time gas mass, the SFR, and the estimated total mass of
IC10.

In spite of the large number of parameters, the number of
observational constraints reproduced by our model makes us confident
in concluding the following:

\begin{itemize}
\item The galaxy IC10 probably formed by means of a slow
gas accretion process with a long infall timescale of the order of 8
Gyr.

\item We do not know a priori the SFH of IC10 but our numerical
simulations suggest that it should have experienced between a
minimum of 4 and a maximum of 10 bursts of star formation. If we
were to trust the present-time SFR measured from FIR and 1.4 GHz,
the duration of the bursts should be longer than the interburst
period, corresponding to a gasping mode of star formation, and the
SFE should not be higher than 0.4 Gyr$^{-1}$. On the other hand, if
we were to assume the present-time SFR inferred from H$\alpha$,
which is higher than the other measurements, then bursts of higher
SFE ($\sim 1$ Gyr$^{-1}$) but longer interburst periods can also be
acceptable. In any case, we exclude a regime of continuous SF.

\item We exclude evolution without galactic winds.
In this case, our model can reproduce neither the present-time gas
fraction nor the metallicity of IC10 at the same time without
including winds. Therefore, galactic winds must occur.

\item A metal-enhanced wind is more likely than a normal
one, otherwise an unreasonable amount of mass is lost by the galaxy,
making it impossible to fit the present-time amount of gas. In our
optimal models (e.g. models mw8, mw9, and mw12), a continuous wind
develops following the first burst that carries away mostly metals,
as assumed in hydrodynamical simulations. Both gasping and bursting
models with metal-enhanced winds can reproduce the macroscopic
properties of IC10 and the scatter in the abundance ratios.

\end{itemize}

\begin{acknowledgements}
J.Y. thanks the hospitality of the Department of Physics of the University of
Trieste where this work was accomplished. J.Y., F.M. and L.M. acknowledge
financial support from PRIN2007 from Italian Ministry of Research, Prot. no.
2007JJC53X-001. J.Y. also thanks the financial support from the
National Science Foundation of China No.10573028, the Key Project
No.10833005, the Group Innovation Project No.10821302, and 973
program No. 2007CB815402.
\end{acknowledgements}


\begin{thebibliography}{}
\bibitem[2005]{asplund05} Asplund, M., Grevesse, N., \& Sauval, A. J. 2005, ASPC, 336, 25
\bibitem[2003]{bell03} Bell, E. F. 2003, ApJ, 586, 794
\bibitem[2000]{borissova00} Borissova, J., Georgiev, L., Rosado, M., Kurtev, R., Bullejos, A., \& Valdez-Guti\'errez, M. 2000, A\&A, 363, 130
\bibitem[1998]{bradamante98} Bradamante, F., Matteucci, F., \& D'Ercole, A. 1998, A\&A 337, 338
\bibitem[2002]{carigi02} Carigi L., Hernandez X., \& Gilmore G. 2002, MNRAS, 334, 117
\bibitem[2003]{crowther03} Crowther, P. A., Drissen, L., Abbott, J. B., Royer, P., \& Smartt, S. J. 2003, A\&A, 404, 483
\bibitem[2004]{demers04} Demers, S., Battinelli, P., \& Letarte, B. 2004, A\&A 424, 125
\bibitem[1999]{dercole99} D'Ercole, A., \& Brighenti, F., 1999, MNRAS, 309, 941
\bibitem[1997]{ellis97} Ellis, R.~S.\ 1997, \araa, 35, 389
\bibitem[2003]{fujita03} Fujita, A., Martin C. L., Mac Low, M.-M., \& Abel, T. 2003, ApJ, 599, 50
\bibitem[1990]{garnett90} Garnett, D. R. 1990, ApJ, 363, 142
\bibitem[2002]{garnett02} Garnett, D. R. 2002, ApJ, 581, 1019
\bibitem[2009]{gavilan09} Gavil\'an, M., Molla, M., \& Diaz, A. I. 2009, arXiv0903.0932
\bibitem[2003]{gil03} Gil de Paz, A., Madore, B. F., \& Pevunova, O. 2003, ApJS, 147, 29
\bibitem[2001]{hidalgo01} Hidalgo-G\'amez, A. M., Masegosa, J., \& Olofsson, K. 2001, A\&A, 369, 797
\bibitem[1995]{hodge95} Hodge, P., \& Miller., B. W. 1995, ApJ, 451, 176
\bibitem[1990]{hodge90} Hodge, P., \& Lee, M. G. 1990, PASP, 102, 26
\bibitem[1936]{hubble36} Hubble, E. 1936, {\em The Realm of the Nebulae}, New Haven: Yale University Press
\bibitem[1979]{huchtmeier79} Huchtmeier, W. K. 1979, A\&A, 75, 170
\bibitem[1988]{huchtmeier88} Huchtmeier, W. K., \& Ritcher, O.-G. 1988, A\&A, 203, 237
\bibitem[2003]{jarrett03} Jarrett, T. H., Chester, T., Cutri, R., Schneider, S. E., \& Huchra, J. P. 2003, AJ, 125, 525
\bibitem[1994]{kennicutt94} Kennicutt, R. C., Jr., Tamblyn, P., \& Congdon, C. E. 1994, ApJ, 435, 22
\bibitem[2009]{kim09} Kim, M., Kim, E., Hwang, N., Lee, M. G., Im, M., Karoji, H., Noumaru, J., \& Tanaka, I. 2009, ApJ, 703, 816
\bibitem[2008]{kniazev08} Kniazev, A. Y., Pustilnik, S. A., \& Zucker, D. B. 2008, MNRAS, 384, 1045
\bibitem[1997]{kobulnicky97} Kobulnicky, H. A., Skillman, E. D., Roy, J.-R., Walsh, J. R., \& Rosa, M. R. 1997, ApJ, 477, 679
\bibitem[2003]{lanfranchi03} Lanfranchi G. A., \& Matteucci F. 2003, MNRAS, 345, 71
\bibitem[1979]{lequeux79} Lequeux, J., Peimbert, M., Rayo, J. F., Serrano, A., \& Torres-Peimbert, S. 1979, A\&A, 80, 155
\bibitem[2006]{leroy06} Leroy, A., Bolatto, A., Walter, F., \& Blitz, L. 2006, ApJ, 643, 825
\bibitem[2009]{loz09} Lozinskaya, T. A., Egorov, O. V., Moiseev, A. V., \& Bizyaev, D. V. 2009, AstL, 35, 730
\bibitem[1999]{mac99} Mac Low, M.-M., \& Ferrara, A. 1999, ApJ, 513, 142
\bibitem[2003]{magrini03} Magrini, L., Corradi, R. L. M., Greimel, R., Leisy, P., Lennon, D. J., Mampaso, A., Perinotto, M., Pollacco, D. L., et al. 2003, A\&A, 407, 51
\bibitem[2009]{magrini09} Magrini, L. \& Gon\c calves, D. R. 2009, MNRAS, 398, 280 (MG09)
\bibitem[1992]{massey92} Massey, P., Armandroff, T. E., \& Conti, P. S. 1992, AJ, 103, 1159
\bibitem[1995]{massey95} Massey, P., \& Armandroff, T. E. 1995, AJ, 109, 2470
\bibitem[2003]{massey03} Massey, P., Olsen K. A. G., \& Parker, J. W. 2003, PASP, 115, 1265
\bibitem[1998]{mateo98} Mateo M. L. 1998, ARA\&A, 36, 435
\bibitem[1994]{melisse94} Melisse, J. P. M., \& Israel, F. P. 1994, A\&AS, 103, 391
\bibitem[2005]{molla05} Moll\'a M., \& D\'\i az A. I., 2005, MNRAS, 358, 521
\bibitem[1996]{molla96} Moll\'a M., Ferrini F., \& D\'\i az A. I. 1996, ApJ, 466, 668
\bibitem[2001]{recchi01} Recchi, S., Matteucci, F., \& D'Ercole, A. 2001, MNRAS, 322, 800
\bibitem[2008]{recchi08} Recchi, S., Spitoni, E., Matteucci, F., \& Lanfranchi, G. A. 2008, A\&A, 489, 555
\bibitem[2001]{richer01} Richer, M. G., Bullejos A., Borissova J., McCall, M. L., Lee, H., Kurtev, R., Georgiev, L., Kingsburgh, R. L., et al. 2001, A\&A, 370, 34
\bibitem[2006]{romano06} Romano, D., Tosi, M., \& Matteucci, F. 2006, MNRAS, 365, 759
\bibitem[1995]{roy95} Roy, J.-R., \& Kunth, D. 1995, A\&A, 294, 432
\bibitem[1999]{sakai99} Sakai, S., Madore, B. F., \& Freedman, W. L. 1999, ApJ, 511, 671
\bibitem[1955]{salpeter55} Salpeter, E. E. 1955, ApJ, 121, 161
\bibitem[2008]{sanna08} Sanna, N., Bono, G., Stetson, P. B., Monelli, M., Pietrinferni, A., Drozdovsky, I., Caputo, F., Cassisi, S., et al. 2008, ApJ, 688, L69
\bibitem[2009]{sanna09} Sanna, N., Bono, G., Stetson, P. B., Pietrinferni, A., Monelli, M., Cassisi, S., Buonanno, R., Sabbi, E., et al. 2009, ApJ, 699, L84
\bibitem[1989]{shostak89} Shostak, G. S., \& Skillman, E. D. 1989, A\&A, 214, 33
\bibitem[1989]{skillman89} Skillman, E.~D., Kennicutt, R.~C., \& Hodge, P.~W.\ 1989, ApJ, 347, 875
\bibitem[1990]{thronson90} Thronson, H. A., Hunter, D. A., Casey S., \& Harper, D. A. 1990, ApJ, 355, 94
\bibitem[2007]{vaduvescu07} Vaduvescu, O., McCall, M. L., \& Richer, M. G. 2007, AJ, 134, 604
\bibitem[2000]{vdb00} van den Bergh, S. 2000, The galaxies of the Local Group (Cambridge:
Cambridge University Press)
\bibitem[1997]{van97} van den Hoek, L. B., \& Groenewegen, M. A. T. 1997, A\&AS, 123, 305
\bibitem[1992]{white92} White, R. L., \& Becker, R. H. 1992, ApJS, 79, 331
\bibitem[1991]{white91} White, S.~D.~M., \& Frenk, C.~S.\ 1991, ApJ, 379, 52
\bibitem[1998]{wilcots98} Wilcots, E. M., \& Miller, B. W., 1998, AJ, 116, 2363
\bibitem[1996]{wilson96} Wilson, C. D., Welch, D. L., Reid, I. N., Saha, A., \& Hoessel, J. 1996, AJ, 111, 1106
\bibitem[1995]{woosley95} Woosley, S. E., \& Weaver, T. A., 1995, ApJS, 101, 181
\bibitem[2010]{yin10} Yin, J., Matteucci, F., \& Vladilo, G. 2010, submitted to A\&A
%
\end{thebibliography}
\end{document}